\documentclass{moriond}

\usepackage{amssymb}
\usepackage{amsmath}
\usepackage{floatrow}

\newcommand{\Photo}{}

\begin{document}
\vspace*{4cm}
\title{The Inverse Square Law And Newtonian Dynamics space explorer (ISLAND)}

\author{ J. BERG\'E }

\address{ONERA -- The French Aerospace Lab,
29 avenue de la Division Leclerc, 
92320 Ch\^atillon, France}

\maketitle\abstracts{
The ISLAND (Inverse Square Law And Newtonian Dynamics) Space Explorer is a new concept to test the gravitational Inverse Square Law at: (1)  submillimeter scale and (2) at the largest Solar System scales (dozens of Astronomical Units --AU).
The main idea is to embark a torsion pendulum at the center of gravity of a dedicated, possibly drag-free and attitude-controlled, interplanetary probe whose gravitational environment is accurately probed by, and corrected for thanks to six ultrasensitive accelerometers arranged as a cross around the torsion pendulum.}

\section{Introduction}

In its century of existence, General Relativity (GR) has been extremely successful at describing gravity. From Mercury's perihelion to gravitational lensing, to gravitational waves indirect and, recently, direct detection, it has passed all experimental tests. Yet, gravitation remains puzzling, as it poses unresolved challenges: on the one hand, observations on cosmological scales cannot be explained by the adjunction of GR to the standard model of particle physics; on the other hand, it is difficult to unify with other known interactions.

Zwicky \cite{zwicky33} was the first to understand the problem of missing matter: galaxies rotate faster than expected based on their observed luminosity. The ``dark matter" conundrum have puzzled astronomers for eighty years, and yet, despite intensive direct and indirect searches in the last decade, no new particle that could account for dark matter has been discovered. Alas, dark matter cannot be explained by a simple observational bias, since it is needed to explain how large scale structures formed in the early Universe: it is the second most significative component of the $\Lambda$CDM cosmological model, accounting for 25\% of the mass-energy budget of the Universe.

Moreover, GR is usually considered to be the ``correct" theory of gravitation on cosmological scales (i.e., at low energy). The $\Lambda$CDM model has been built from this assumption, and from the observation of both dark matter and of the accelerated expansion of the Universe \cite{riess98;perlmutter99}. This unexpected behavior may be explained by the presence of some dynamical, repulsive, ``dark energy". This new component (which accounts for 70\% of the mass-energy budget of the Universe) could be an evolving scalar field (quintessence), or more simply Einstein's cosmological constant $\Lambda$, which may be linked to the vacuum energy. However, at least two problems are encountered: (1) the {\it cosmological constant problem} arises from the fact that the observed value of $\Lambda$ is around 120 orders of magnitude smaller than the naive expectation from Quantum Field Theory (QFT), that should be of the Planck mass, and (2) the {\it concordance problem} arises from the fact that the current energy density of dark energy is of the same order as that of dark matter, which is surprising since dark matter and dark energy are separate fields that should evolve independently.

GR (and by extension Newtonian gravity) describes low energy phenomena, but does not take into account the quantum nature of matter, and therefore fails at describing gravity for high-energy phenomena, that are well described by QFT. On the opposite, QFT and its extensions assume that spacetime is flat, whereas GR's spacetime is intrinsically curved. 
It is not clear, neither, why gravity is so weak compared with the other interactions (it is $10^{32}$ times weaker than the weak force), nor how to unify it with them.

GR describes the gravitational force as mediated by a single rank-2 tensor field. There is good reason to couple matter fields to gravity in this way, but there is no good reason to think that the field equation of gravity should not contain other fields. The simplest way to go beyond GR and modify gravity is then to add an extra scalar field: such scalar-tensor theories are well established and studied theories of Modified Gravity (MG). From a phenomenological point of view, they link the cosmic acceleration to a deviation from GR on large scales. They can therefore be seen as candidates to explain the accelerated rate of expansion without the need to consider dark energy as a physical component. Furthermore, they arise naturally as the dimensionally reduced effective theories of higher dimensional theories, such as string theory; hence, testing them can allow us to shed light on the low-energy limit of quantum gravity theories.

Scalar fields that mediate a long range force able to affect the Universe's dynamics should also significantly modify gravity in the Solar System, in such a way that GR should not have passed any experimental test. Screening mechanisms have been proposed to alleviate this difficulty \cite{screening}. In these scenarios, (modified) gravity is environment-dependent, in such a way that gravity is modified at large scales (low density) but is consistent with the current constraints on GR at small scale (high density). 

All theoretical attempts at explaining the three limitations mentioned above (string theory, MG, MOND) modify GR and Newton dynamics, either at small scale or very large scales, or both. In particular, if one of them is correct, we should detect a violation of the gravitational Inverse Square Law (ISL). Hence, in the weak field limit, measurements of the dynamics of gravitationally bound objects (e.g. the behavior of a torsion balance in the Earth gravity field, the trajectory of an interplanetary probe, or the receding of galaxies) should show a deviation from what is expected from Newton's equations.

We introduce the Inverse Square Law And Newtonian Dynamics space explorer (ISLAND) to look for such deviations on the submilliter and Solar System scales.

\section{Tests of the inverse square law at small and large scales}

\begin{figure}[t]
\includegraphics[width=0.47\textwidth]{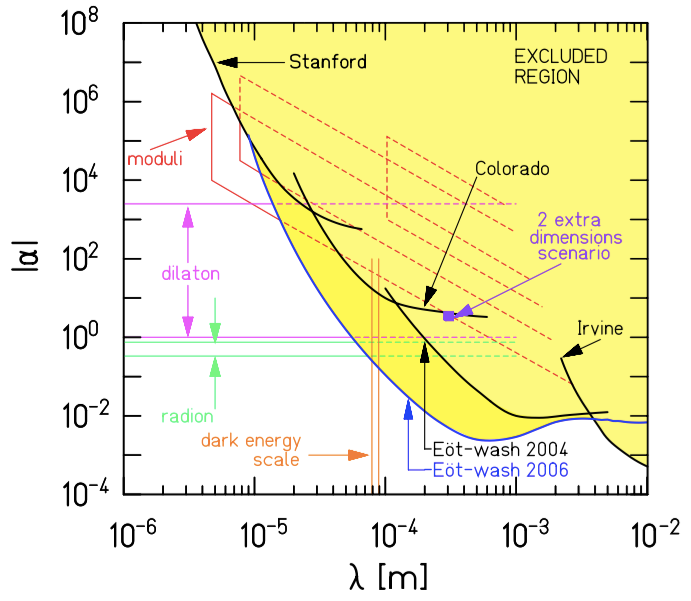}
\includegraphics[width=0.45\textwidth]{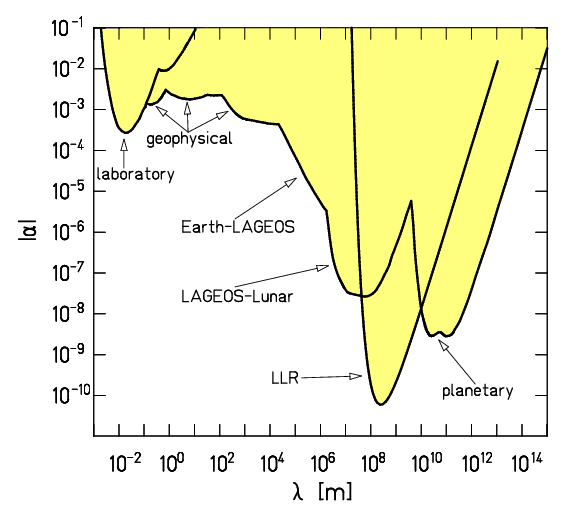}
\caption{95\% CL constraints on ISL violating Yukawa interactions. Left panel: $\lambda<1$cm. Right panel: $\lambda>1$cm}
\label{fig_yukawa_constraints}       
\end{figure}

Deviations from the ISL are usually parametrized with a Yukawa potential. The gravitational potential created by a point-mass of mass $M$ at a distance $r$ is then given by

\begin{equation}
V(r) = -\frac{GM}{r} \left(1 + \alpha e^{-r/\lambda} \right),
\end{equation}
where $G$ is the gravitational constant, $\alpha$ is the (dimensionless) strength of the Yukawa potential relative to Newtonian gravity, and $\lambda$ is its range. Fig. \ref{fig_yukawa_constraints} shows the current constraints on ISL-violations from a Yukawa potential. 
The left panel focuses on subcentimeter experiments. .
The right panel focuses on scales larger than the centimeter; the best constraints come from the measurement of the Moon orbit with the Lunar Laser Range and from planetary motion.

Adelberger et al \cite{adelberger03} provide a thorough review of experimental tests of small scale gravitation.
Most notably, the E\"ot-Wash group of the University of Washington uses low-frequency torsion pendulums to test the ISL \cite{ew}. 
Their pendulum consists in two parallel disks with regularly placed cylindrical missing parts, separated by a given gap: the upper disk (the detector) hangs from a thin wire and is let free to oscillate when the lower disk (the attractor) rotates. They then use an optical device to measure the deviation of the detector from its equilibrium position, and relate it to the torque created by the rotation of the attractor. 
They could provide strong constraints on the Yukawa ($\alpha$, $\lambda$) plane \cite{ew_yukawa}, as shown by the ``E\"ot-Wash'' curves in the left panel of Fig. \ref{fig_yukawa_constraints}.
It is clear that a large part of the ($\alpha$, $\lambda$) plane at submilliter scale is already excluded, with a strong degeneracy between 95\% confidence level of the allowed strength $\alpha$ and range $\lambda$. The allowed region still gives possibilities for new physics, but improving the constraints on $\lambda \leqslant 10 \mu$m, with a sensitivity on $\alpha$ similar to that of E\"ot-Wash, may provide significant constraints on string-inspired theories.

Gravity at cosmogical scales can be tested through observations (eg. weak gravitational lensing or baryon acoustic oscillations). Surveys like Euclid will most probably bring tight constraints on dark matter and dark energy.

Solar-System-scale observations and/or experiments can allow us to fill the gap between submilliter- and laboratory-scale measurements, and cosmological constraints. The tightest constraints come from measurements by the Lunar Laser Range (LLR \cite{murphy13}). 
Larger scales can be probed through planetary ranging and ephemeris \cite{fienga}. 
Such techniques provide a tight constraint on $\alpha$; however, this constraint becomes quickly loose as the scale increases towards the outer solar system scale.
Probing those scales can be done by measuring the orbit of an outbound spacecraft as it drifts away from the Sun and the planets. The most notable test was performed by NASA during the extended Pioneer 10 \& 11 missions. The test resulted in the so-called Pioneer anomaly, finally accounted for by an anisotropic heat emission from the spacecrafts themselves \cite{pioneer}. However, with no direct measurement of the non-gravitational forces acting on the Pioneer 10 \& 11 spacecrafts, one could still argue that the case is not completely closed.

The effect of non-gravitational forces could be definitely accounted for if we were able to measure them directly, instead of relying on spacecraft and environment models. This can be easily done with an onboard accelerometer, as proved by missions like LISA Pathfinder \cite{armano16} or MICROSCOPE \cite{berge15}. 
Combining radio tracking data with the accelerometer's  direct and model-independent measurements of non-gravitational accelerations, it becomes possible to improve the comparison between the estimated spacecraft's gravitational acceleration and theory by orders of magnitude. 
The Outer Solar System (OSS) mission \cite{christophe12} investigated the opportunity to fly an accelerometer in deep space. 
More recently, it was shown that we could improve the constraints of Fig. \ref{fig_yukawa_constraints} by at least two orders of magnitude, if we are able to measure the orbit with a 1 meter accuracy, and to measure an absolute acceleration of order $10^{-17}$ m/s$^2$, albeit with a mission plan different from that of ISLAND \cite{buscaino15}.

\section{ISLAND}

ISLAND aims to improve experimental constraints on those two scales that are currently loosely constrained: under the millimiter, and at the largest Solar System scales.
On submilliter scales, improving the constraints on the Yukawa strength about $\lambda \approx 10\mu$m by two orders of magnitude would allow us to constrain string-theory inspired models.
Flying ISLAND up to 100 AU would then allow us to improve the constraints on the Yukawa strength for $10^{12} {\rm m} \leqslant \lambda \leqslant 10^{13} {\rm m}$ by at least two orders of magnitude.

\subsection{Need to go to space (for submilliter-scale tests)}

Current on-ground torsion pendulum experiments are mostly limited by thermal noise from the wire, by the difficulty to align the plates with the Earth gravity field, and by seismic noise. In particular, seismic noise limits the smallest separation between the attractor and detector to a few dozens microns \cite{kapner07}, which in turn limits the region of the Yukawa parameters plane that can be probed. 
Those difficulties are alleviated when performing the experiment in space, in a drag-free probe, with an electrostatic torsion pendulum at the center of gravity of the spacecraft: there is no wire anymore, there is no gravity field to which the apparatus should be aligned, and no seismic noise, allowing us to probe smaller scales. Furthermore, if the torsion pendulum lies at the center of a gradiometer, all remaining gravity gradients (e.g. from the Sun gravitational field) can be measured and corrected for.
Finally, a space environment is needed to optimize the electrostatic levitation and control of the torsion pendulum's detector.

Going to space will be necessary to experimentally investigate screening mechanisms. Indeed, environment-dependent screenings can prevent an experiment from detecting any deviation from the ISL on the Earth, while allowing the same experiment to detect a deviation in a space environment. Similarly, testing for screening mechanisms at different distances from the Sun might improve our understanding on their environment-dependence.

\subsection{Measurement concept} \label{sect_concept}

Although both tests can be done independently, the small scale test takes advantage of being done at the center of mass of a gradiometer onboard a drag-free spacecraft. The gradiometer allows us to control the spacecraft's drag-free and to perform the small scale test on an optimal local gravitational environment. In turn, the gradiometer's accelerometers measure the non-gravitational accelerations required by the large scale test. Both tests thus share the same hardware.

The test of the ISL at submilliter scales will be performed with an electrostatic torsion pendulum, made of two parallel plates, based on the E\"ot-Wash pendulum. The pendulum will be set at the center of gravity of a gradiometer (whose position will coincide with the center of mass of the spacecraft) in order to take advantage of an optimally clean gravitational environment.

The test of the ISL at large scale will be performed by precisely tracking the orbit of the spacecraft as it cruises the Sun's gravitational field. The gradiometer's accelerometers will allow us to measure and correct for non-gravitational forces, in a model-independent way.

\subsection{Submilliter-scale test: torsion pendulum}

ISLAND will perform submilliter-scale tests with an electrostatic torsion pendulum based on the E\"ot-Wash concept \cite{adelberger03}. The E\"ot-Wash torsion pendulums consist in two parallel disks with regularly placed cylindrical missing parts, separated by a given gap: the upper disk (the detector) hangs from a thin wire and is let free to oscillate when the lower disk (the attractor) rotates. An optical device is used to measure the deviation of the detector from its equilibrium position, and relate it to the torque created by the rotation of the attractor. The interaction between the two disks is gravitational (electromagnetic forces are minimized), which allow them to set constraints on a possible deviation from the gravitational ISL.

\begin{figure}[t]
\includegraphics[width=0.47\textwidth]{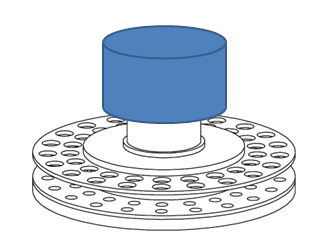}
\includegraphics[width=0.47\textwidth]{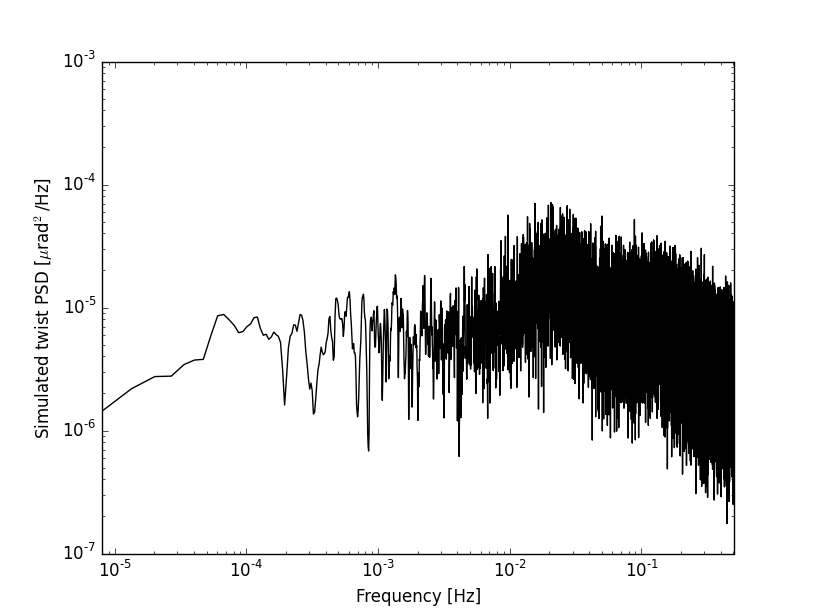}
\caption{Left: Torsion pendulum concept; a lower disk rotates and applies a torque to the upper disk (detector); its deviation from its equilibrium position is measured through a capacitive device (blue); the capacitive device also permits the levitation of the detector, thereby acting as a wire in a gravity field. Right: expected twist noise level.}
\label{fig_pendulum}       
\end{figure}

As the ISLAND torsion pendulum is in microgravity onboard the spacecraft, there is no preferred direction for a disk to hang from a wire. That is why we plan to replace the wire by a capacitive system able to control the detector and measure its deviation (Fig. \ref{fig_pendulum}). ONERA has already built a prototype \cite{willemenot00} of an electrostatic torsion pendulum, thereby proving that it can be possible to replace the wire by a capacitive system.
As shown in Fig. \ref{fig_pendulum}, the envisioned capacitive system would consist of an inner cylinder controled by electrodes glued to an external cylinder (blue). The pendulum detector would be attached to the inner cylinder, and hence its motion would be directly measured by the capacitive system through the deviation of the inner cylinder from its equilibrium position.

This geometry is reminiscent of the cylindrical differential electrostatic accelerometers onboard the MICROSCOPE mission. The level of their noise up to $10^{-2}$ Hz is of the order of $10^{-5}$ $\mu$rad$^2$/Hz, almost one order of magnitude lower than the noise of the E\"ot-Wash pendulum at the frequency of their test (right panel of Fig. \ref{fig_pendulum}) . Hence, with a technology adapted from that of MICROSCOPE, we should be able to reach the desired accuracy in twist measurement.

Given this expected noise level, we should perform the test of the ISL at a frequency ranging from $10^{-4}$ Hz to $10^{-2}$ Hz. As we will show below, this is in agreement with the bandwith of the GAP accelerometer, needed to clean the gravitational environment of the torsion pendulum.
Assuming that the attractor and detector are made up of $N \approx 10$ holes, the attractor rotation rate $\omega$ should then be fixed between $10^{-5}$ Hz and $10^{-3}$ Hz to ensure that $N\omega$, the frequency of the expected Yukawa signal, lies in the bandwidth of the accelerometers. 

The ISLAND spacecraft should be able to be in a drag-free and fine attitude-control mode when the pendulum will be in function. As a consequence, all non-gravitational backgrounds will be canceled at the position of the pendulum. 
Therefore, in the spacecraft reference frame, where the test will be performed, only the spacecraft self-gravity will then contribute, but it should be easy to correct for it, either by design of the spacecraft or {\it a posteriori} data analysis correction.

\subsection{Large-scale test: orbit determination}

To test gravity at large scales, ISLAND will rely on the accurate determination of its trajectory, as it cruises the Solar System and is affected by gravitational forces from the Sun, planets, moons and other bodies, as well as relativistic effects.
ISLAND's orbit can be determined through ranging, Doppler tracking and Very Long Baseline Interferometry (VLBI).

The ISLAND mission will use modern radio-techniques that have been developed for deep space planetary missions. The availability of range observables will significantly reduce the uncertainty in the determination of the heliocentric distances of the probe. Meanwhile the VLBI techniques give new angular observables related to transverse motion with respect to the line of sight. The combination of measurements based on these radio techniques and accelerometry will allow ISLAND to measure gravity in the Solar System with a greatly improved measurement accuracy and better control of the systematics. As mentioned by Buscaino et al \cite{buscaino15}, a one-meter precision in the spacecraft ranging is attainable by ranging measurement with the Deep Space Network (DSN). Such a precision would significantly improve the constraints on a Yukawa deviation at scales higher than 10 AU.

With the presence of the accelerometer, the navigation of the probe is no longer affected by uncertainties or inaccuracies in the models of the non-gravitational acceleration. This approach has been used for geodesy missions such as CHAMP, GRACE and GOCE, as well as MICROSCOPE \cite{touboul}. The additional accelerometry observable removes parameters to be fitted in the orbit determination process and consequently improves its quality. It also measures the temporal fluctuations of the non-gravitational acceleration, which are not taken into account by models. Finally, it removes the correlations that appear in the orbit determination process between the non-gravitational acceleration and the gravitational acceleration, when the former are not measured.

\subsection{Gradiometer and accelerometers}

We want to be able to accurately measure the gravity gradients around ISLAND's torsion pendulum. One way to achieve it is to design a gradiometer with six accelerometers, that should be placed in a cross figure, like ESA's GOCE's gradiometer. Placing the torsion pendulum at the center of gravity of such a gradiometer will allow us to efficiently estimate any gravity gradient to which it is subjected. This knowledge will allow us to correct for gravity gradients at the data analysis stage.

A gradiometer can use biased accelerometers, since it is possible to correct for the bias between accelerometers, and because what is important for a gradiometer is relative accelerations. The situation would be identical, for a drag-free spacecraft, when tracking its trajectory: a drag-free system works well with relative acceleration. However, if the spacecraft is not always in a drag-free mode during its cruise phase, correcting for non-gravitational accelerations can be done only if we are able to have unbiased acceleration measurements. Therefore, an unbiased (DC) accelerometer is needed. ISLAND could use six such accelerometers, or only one, that would serve as a reference during non-drag-free periods (between torsion pendulum experiments).

Such a DC accelerometer is currently under development at ONERA: the Gravity Advanced Package (GAP) is composed of an electrostatic accelerometer (MicroSTAR), based on ONERA's expertise in the field of accelerometry and gravimetry (CHAMP, GRACE, GOCE and MICROSCOPE missions), and a bias calibration system. Ready-to-fly technology is used with original improvements aimed at reducing power consumption, size and weight. The bias calibration system consists in a flip mechanism which allows for a 180$^{\rm o}$ rotation of the accelerometer to be carried out at regularly spaced times. The flip allows the calibration of the instrument bias along 2 directions, by comparing the acceleration measurement in the two positions \cite{lenoir}.

The three axes electrostatic accelerometers developed at ONERA are based on the electrostatic levitation of the instrument inertial mass with almost no mechanical contact with the instrument frame. The test-mass is then controlled by electrostatic forces and torques generated by six servo loops applying well measured equal voltages on symmetric electrodes. Measurements of the electrostatic forces and torques provide the six outputs of the accelerometer.

\begin{figure}[t]
\includegraphics[width=0.53\textwidth]{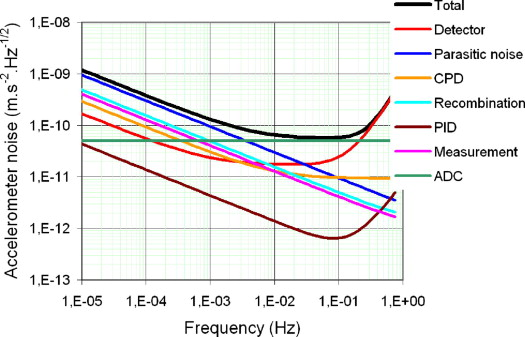}
\includegraphics[width=0.4\textwidth]{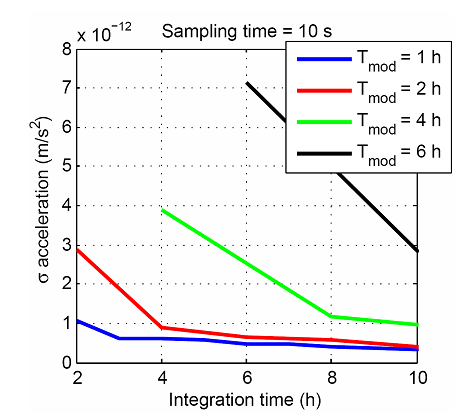}
\caption{Current GAP performance. {\it Left}: MicroSTAR's noise. {\it Right}: Bias rejection system's resolution.}
\label{fig_gap_perfo}       
\end{figure}

MicroSTAR is optimized for measurements in a bandwith of [10$^{-5}$ - 10$^{-1}$] Hz. Assuming that the rotation rate of the torsion pendulum is between $10^{-5}$ Hz and $10^{-3}$ Hz, the signal from the torsion pendulum will be at a frequency well  within MicroSTAR's bandwidth. Furthermore, in order to reach the requirement for the orbit tracking, MicroSTAR's bias must be rejected. 
The left panel of Fig. \ref{fig_gap_perfo} shows the current noise of the MicroSTAR accelerometer, with different contributors: a level of 10$^{-9}$ m/s$^2$/Hz$^{1/2}$ is obtained over [10$^{-5}$ - 1] Hz, with a measurement range of $1.8 \times 10^{-4}$ m/s$^2$. The bias modulation signal, consisting in regular 180$^{\rm o}$ flips of the accelerometer, allows us to band-pass filtering the accelerometer noise around the modulation frequency. The right panel of Fig. \ref{fig_gap_perfo} shows, for different periods of modulation and calibration, that after rejecting the bias, GAP is able to measure absolute accelerations down to $10^{-12}$ m/s$^2$.

\section{Conclusion}

The yet unexplained acceleration of the cosmic expansion, as well as the conflict between general relativity and quantum field theory  at high energy, force us to think beyond, and finely test, established frames. Any deviation from the gravitational inverse square law will be a smoking gun for new physics beyond Einstein, and will shed light towards a viable new theory of gravitation. We propose to look for such a deviation both at submilliter scales with an electrostatic torsion pendulum onboard a dedicated drag-free and attitude-controlled spacecraft; additionally, monitoring the trajectory of the spacecraft and measuring the non-gravitational accelerations that are acted upon it will allow us to test gravity at large scales.

\section*{Acknowledgments}

I acknowledge financial support from the UnivEarthS Labex program at Sorbonne Paris Cit\'e (ANR-10-LABX-0023 and ANR-11- IDEX-0005-02). I want to thank the members of the ISLAND team, from ONERA, IAP, CEA and ZARM, for useful discussions.

\section*{References}


\end{document}